\pdfoutput=1

\documentclass[aps,prd,twocolumn,superscriptaddress,nofootinbib]{revtex4-2}

\usepackage{newtxtext,newtxmath}
\usepackage{abrev-jour-long}
\usepackage[hidelinks]{hyperref}

\usepackage[T1]{fontenc}
\usepackage{ae,aecompl}
\usepackage{bm}

\usepackage{graphicx}	
\usepackage{amsmath}	

\usepackage{color}

\begin{document}
\title[Chaos in the orbital motion of pulsating objects]{
Chaotic orbital dynamics of pulsating objects in dark matter halos
}


\author{Ronaldo S. S. Vieira}
\email{ronaldo.vieira@ufabc.edu.br}
\affiliation{Centro de Ci\^encias Naturais e Humanas, Universidade Federal do ABC, 09210-580 Santo Andr\'e, SP, Brazil}
\author{Ricardo A. Mosna}
\email{mosna@unicamp.br}
\affiliation{Departamento de Matem\'atica Aplicada, Universidade Estadual de Campinas, 13083-859,  Campinas,  S\~ao Paulo,  Brazil}   
\date{\today}


\begin{abstract}
Stellar pulsation is nowadays a widely understood subject. However, there has been no research about the effects of pulsations on the star's orbital dynamics throughout the galaxy. We investigate whether these oscillations can cause chaotic behaviour in the stellar orbit if it is immersed in background dark matter halos. We show that for a certain range of parameters of the pulsating object these effects are present in a general background matter field. However, for the typical range of parameters of realistic astrophysical scenarios, these effects are too small to be detectable.
\end{abstract}

\maketitle



\section{Introduction}

It dates back to the 17th century the observational evidence that some stars have periodic variation in their brightness \cite{carroll2006introduction}. 
Theoretical foundations of this phenomenon, attributing its origin to stellar pulsation, were first developed in the beginnings of the 20th century.
The theory of stellar pulsation is now mature and most of its details are fully understood \cite{cox1974RepPPhys, cox1980TSP}.
In particular, the linear theory of stellar pulsation is well developed, and its nonlinear aspects are widely studied \cite{	takeuti2012nonlinear}, being an active research field nowadays. However, as far as we know, there is no study in the literature regarding the effects of stellar pulsation in the star's orbital dynamics throughout the Galaxy. This is the gap we intend to fill in, at least formally. For reasons explained in the next sections, we will focus on the orbital dynamics of pulsating objects immersed in dark matter halos.

Dark matter halos were first proposed in order to explain the flat or even rising profiles of rotation curves of spiral galaxies for large radii, leading to successful predictions \cite{sofueRubin2001ARAA, einasto2009arXiv, sanders2010darkmatter}.
The observation of dark matter phenomena in other scales, such as in clusters of galaxies (e.g., Ref.~\cite{cloweEtal2006ApJ}) and cosmology (e.g., Ref.~\cite{bartelmann2010RvMP} and references therein) corroborated the dark matter halo hypothesis in spiral and elliptical galaxies.

In this paper we show that pulsating objects may have chaotic orbital motion when immersed in dark matter halo potentials.
This is investigated by applying a recently developed Hamiltonian formulation for the centre-of-mass motion of pulsating extended bodies in the quadrupole approximation \cite{bolinha}. This allows us to analyse the deviations of the extended body's trajectory from its corresponding unperturbed (integrable) test-particle trajectory, and to, in principle, detect chaos due to time-dependent configurations of the body. In order to do that, we analyze the dynamics of extended bodies (as toy models for astrophysical objects) subjected to different dark matter halo fields and argue that, for a certain range of parameters, the appearance of chaos in the orbital dynamics of pulsating objects is a generic phenomenon in the presence of any background matter distribution. Unfortunately the model parameters considered in this paper lie outside the order-of-magnitude estimates for an actual star in a typical galaxy. In this way, this effect may be only formally related to the dynamics of realistic scenarios of pulsating stars in galactic halos.

\begin{figure*}
	\includegraphics[width=0.9\columnwidth]{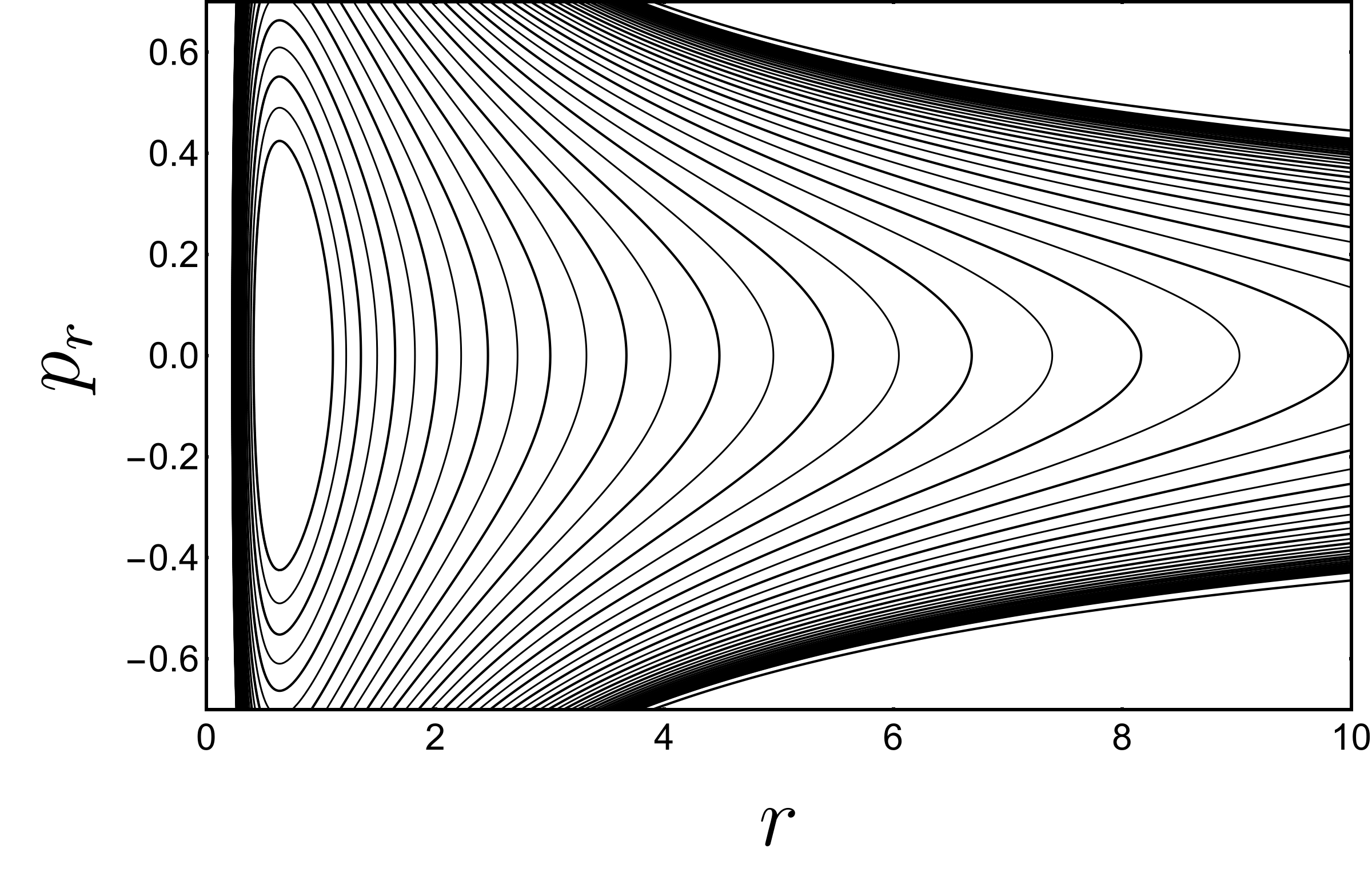}\qquad\qquad
	\includegraphics[width=0.9\columnwidth]{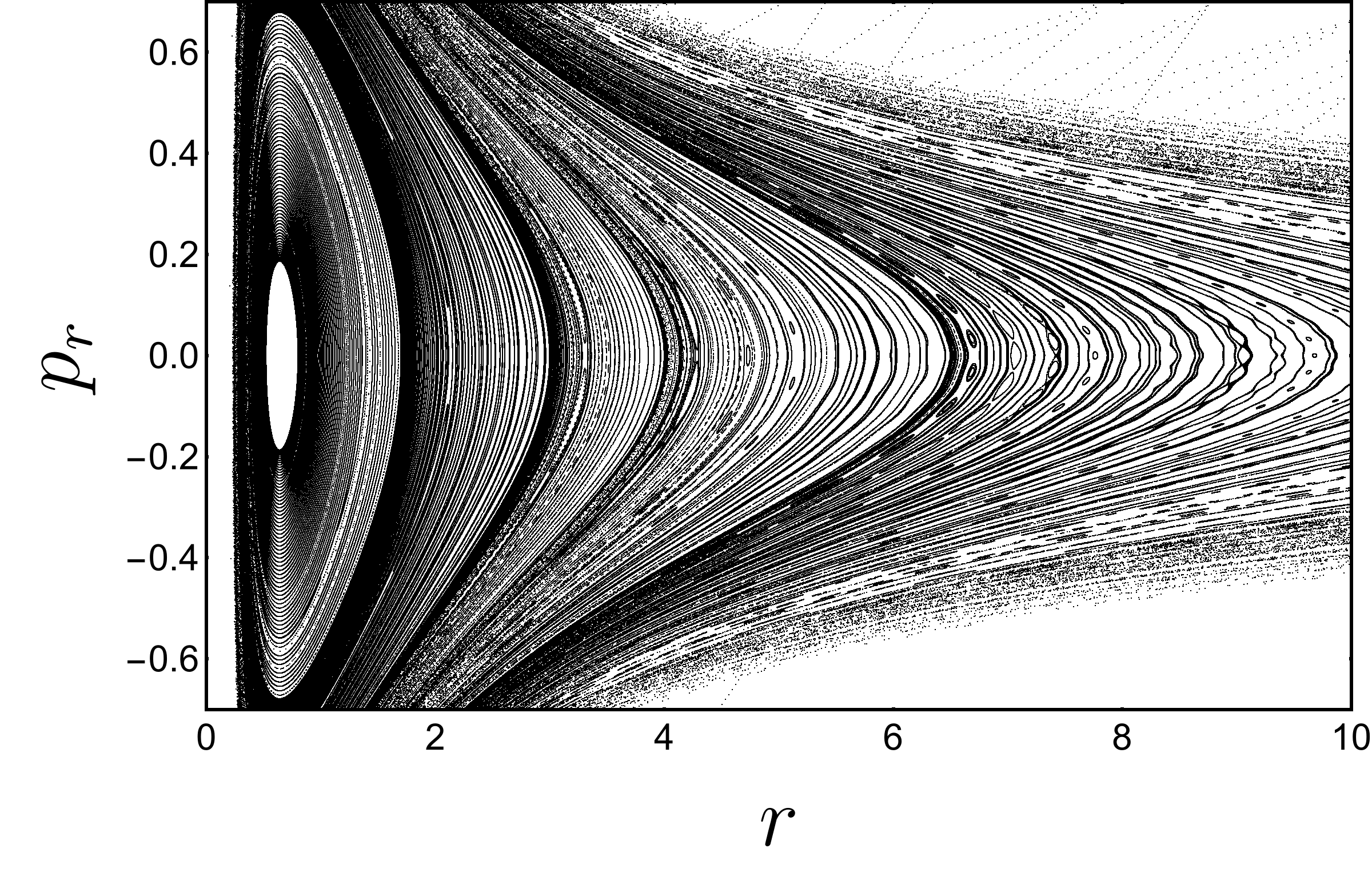}
	\includegraphics[width=0.9\columnwidth]{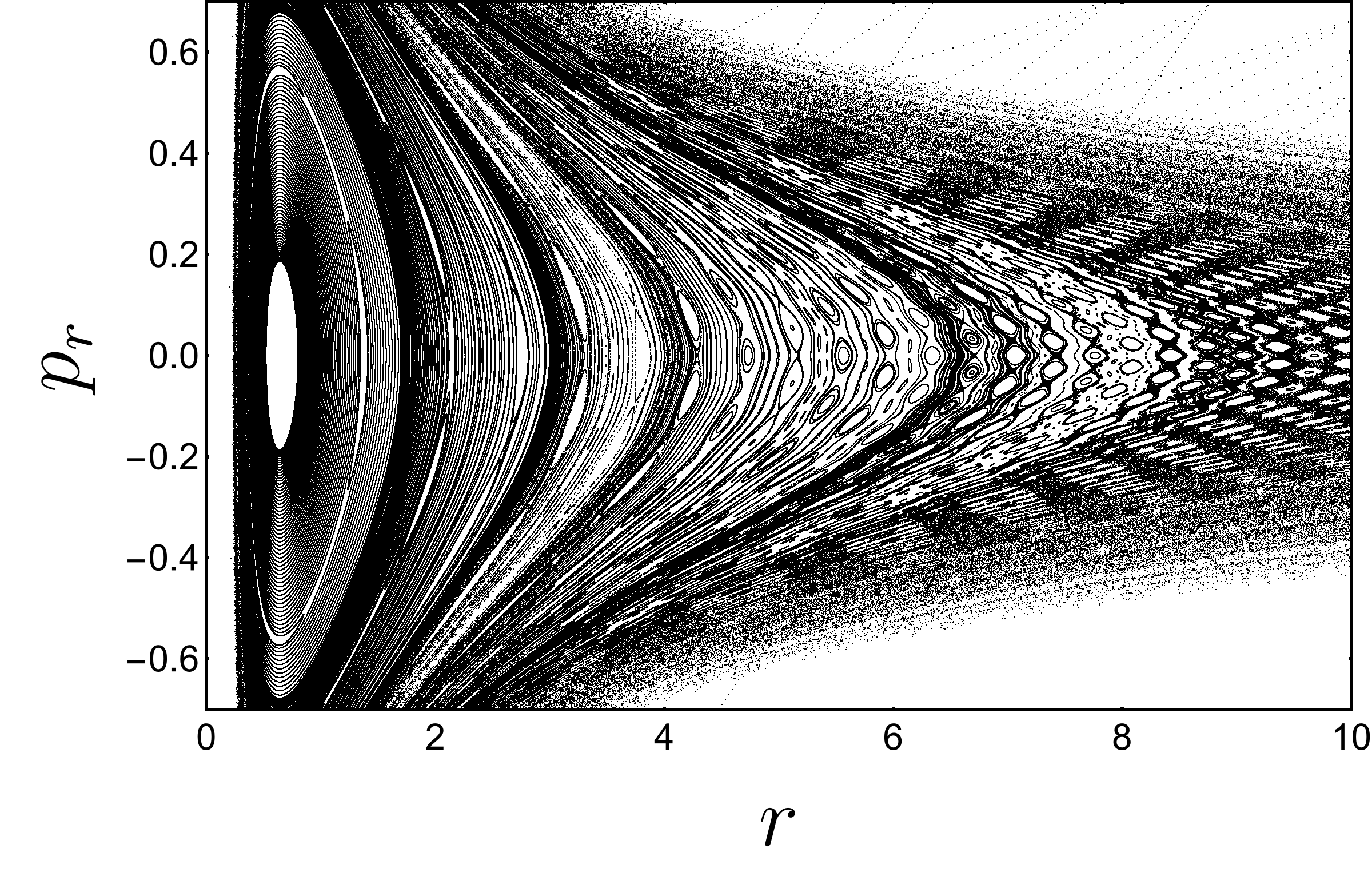}\qquad\qquad
	\includegraphics[width=0.9\columnwidth]{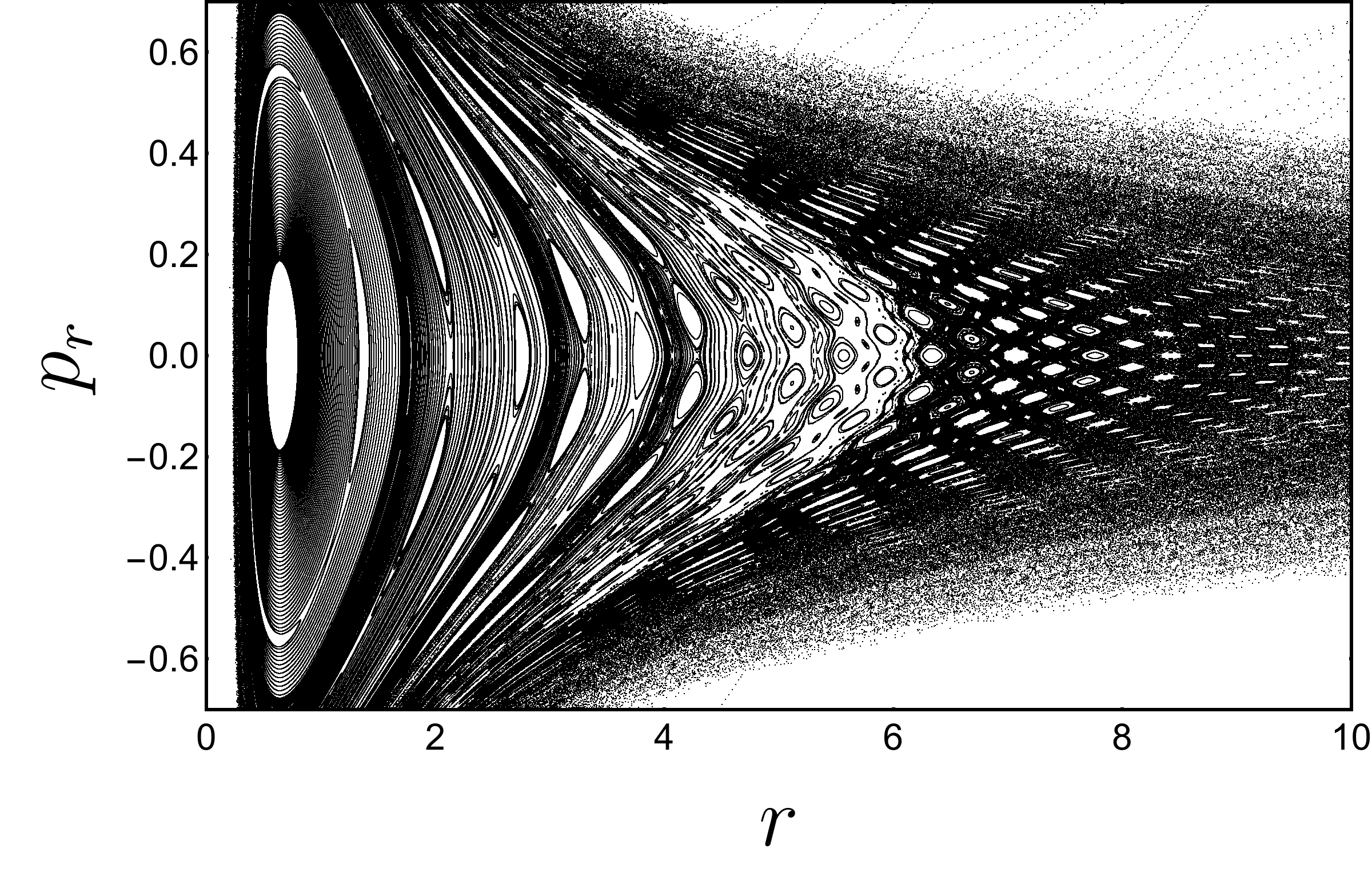}
	\caption{Top, left: phase portrait of a test particle under the Plummer dark halo potential (\ref{eq:PlummerPot}), in units where $GM=1$, $b=1$. We fix $L^2=0.1$. The blank regions correspond to orbits escaping to infinity.
		Top, right: corresponding $(r-p_r)$ stroboscopic Poincar\'e map for the pulsating object orbits, with $q_0/m=0.0001$ and $\Omega=2.0$. 
		Bottom, left: stroboscopic Poincar\'e map for the pulsating object orbits, with $q_0/m=0.0005$ and $\Omega=2.0$. 
		Bottom, right: stroboscopic Poincar\'e map for the pulsating object orbits, with $q_0/m=0.001$ and $\Omega=2.0$.
		The stroboscopic maps clearly show the appearance of resonant chains immersed in a wide chaotic region, which increases as $q_0/m$ gets larger.
	}
	\label{fig:Plummer}
\end{figure*}

\section{Orbital dynamics of spherical test bodies}
\label{sec:dynamics}

We start by briefly reviewing the formalism of \cite{bolinha} as applied to the case of spherically symmetric test bodies.

The total force exerted on an extended test body by a gravitational source with potential $\Phi$ is given by
\begin{equation} \label{eq:force}
{\bm F} = - \int \rho\, \nabla \Phi\, d^3 x\,,
\end{equation}
where $\rho({\bm x, t})$ is the body's mass density distribution. The total torque with respect to the body's centre of mass is ${\pmb \tau} = - \int \rho\,({\bm x} - {\bm z})\,\times\, \nabla \Phi\, d^3 x$. 
The resulting equations of motion are therefore given by $m\, {\ddot{\bm{z}}} = {\bm F}$ and ${ \dot{\bm{S}}} = {\pmb \tau}$, where ${\bm z}={\bm z}(t)$ is the body's centre of mass, and $m$ and $\bm{S}$ are the total mass and intrinsic angular momentum (spin)  of the body.

\subsection{Quadrupole approximation for spherical stars}

Let us assume that the test body is small enough such that the potential gradient may be approximated by a second-order expansion around its centre of mass. The total force on the body can then be written as
\begin{equation} \label{eq:forcequad}
F_i = -m\,\partial_i\,\Phi - \frac{1}{2} Q^{jk}\,\partial_j \partial_k (\partial_i \Phi)\,,
\end{equation}
where the quadrupole tensor $Q^{jk}$ of the mass distribution, with respect to the centre of mass, is given by
\begin{equation}
Q^{jk}(t) = \int \rho({\bm x},t)\,
\Big(x^j - z^j(t)\Big)\Big(x^k - z^k(t)\Big)\,d^3x.
\end{equation}
It is important to note that the equations of motion do not fix the quadrupole dynamics, which may be freely prescribed, for instance, by inner mechanisms controlling the internal mass redistribution of the body \cite{harte2021AcAau}.

A spherically symmetric test body yields $Q^{ij} = q\,\delta^{ij}$, where the \textit{inertia parameter} $q$ is given by \cite{bolinha}
\begin{equation}
q(t) = \frac{1}{3}\int \rho({\bm x},t)\, |{\bm x} - {\bm z}(t)|^2 \,d^3x.
\end{equation}
Thus, for spherical bodies, the inertia parameter $q$ is the sole internal degree of freedom which contributes to the gravitational dynamics at the level of the quadrupole approximation. Moreover, for such a spherically symmetric configuration, the torque $\bm{\tau}$ vanishes by symmetry \cite{harte2021AcAau}.

It follows from equation~(\ref{eq:forcequad}) that $\ddot{\bm{z}} = - \nabla\,\mathcal{V}(\bm{z},\,t)$, where 
\begin{equation}\label{eq:potentialBall}
\mathcal{V}({\bm x},t) = \Phi({\bm x}) + \frac{1}{2 m}\,q(t)\,\nabla^2\,\Phi({\bm x})\,.
\end{equation}
As a result, we end up with a Hamiltonian formulation for the centre-of-mass motion of the body, with Hamiltonian
\begin{equation}\label{eq:testbodyHamiltonian}
\mathcal{H} = \frac{1}{2}\,{\bm p}^2 + \mathcal{V}\,,
\end{equation}
where $\bm p$ is the specific linear momentum of the centre of mass.

\subsection{Quadrupole model for stellar pulsation}

Most variable stars oscillate in their fundamental radial mode, or `breathing mode' \cite{carroll2006introduction}. In this way, instead of building a detailed model for their radial oscillations we consider, for the present purposes, an effective time-periodic oscillation $q(t)$ in the inertia parameter,
\begin{equation}\label{eq:qt}
q(t) = q_0\,\left[1+\cos(\Omega\, t)\right]\,,
\end{equation}
with $q_0>0$. According to this simplified model, the pulsating object oscillates between the point-particle limit ($q=0$) and a configuration with a maximum value $q_{\rm max}=2q_0$.	


\begin{figure*}
	\includegraphics[width=0.9\columnwidth]{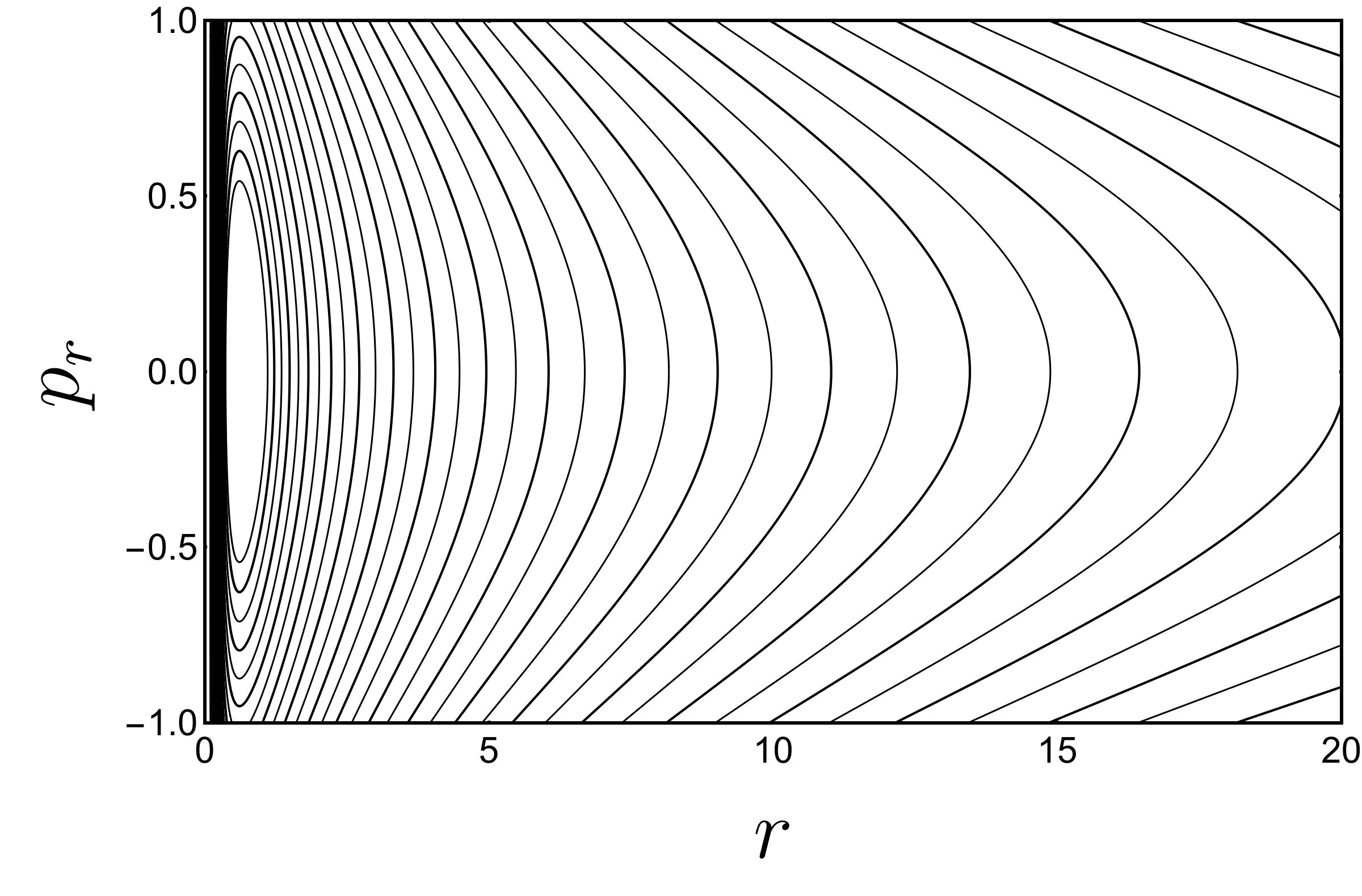}\qquad\qquad
	\includegraphics[width=0.9\columnwidth]{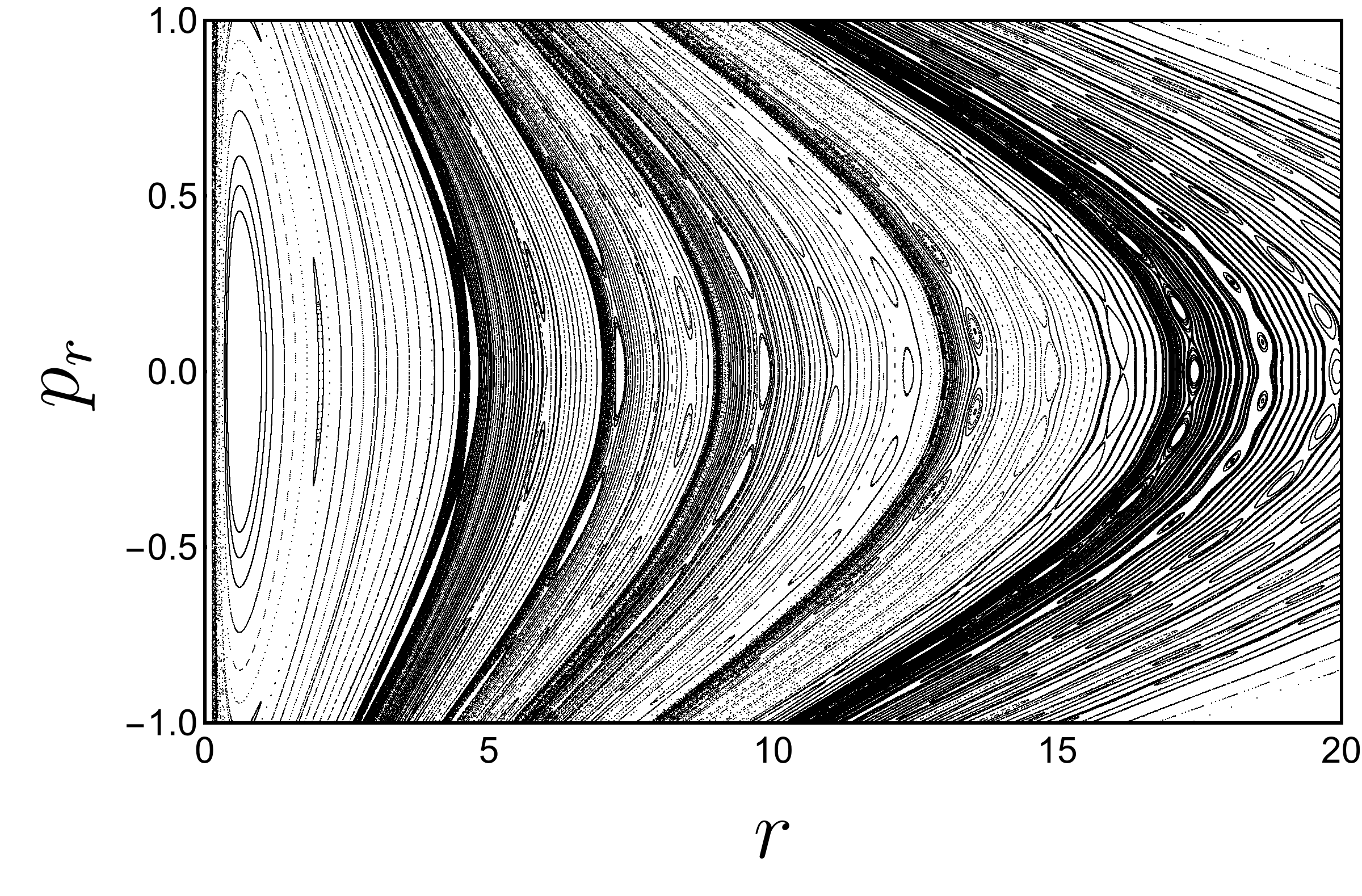}
	\includegraphics[width=0.9\columnwidth]{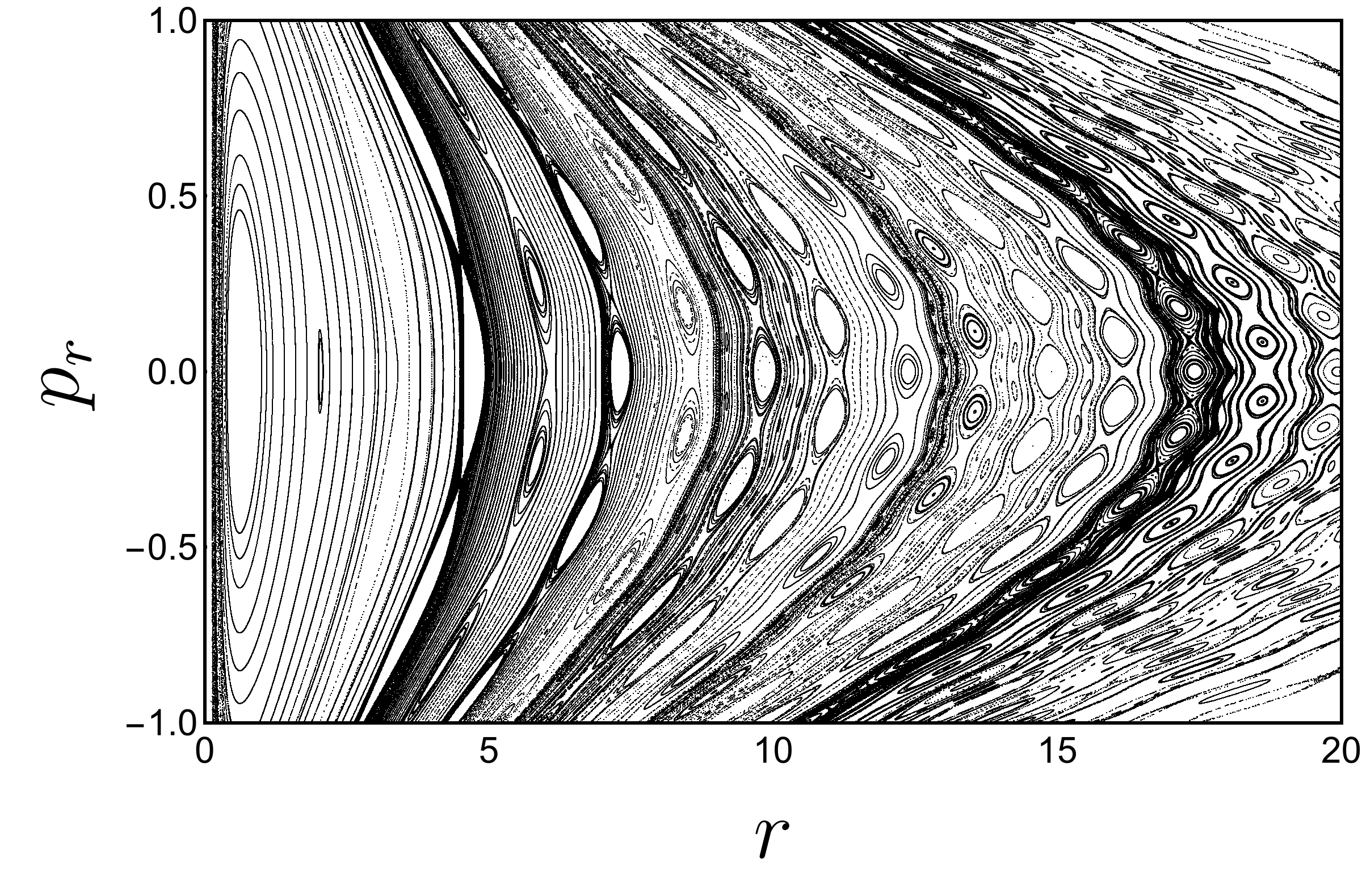}\qquad\qquad
	\includegraphics[width=0.9\columnwidth]{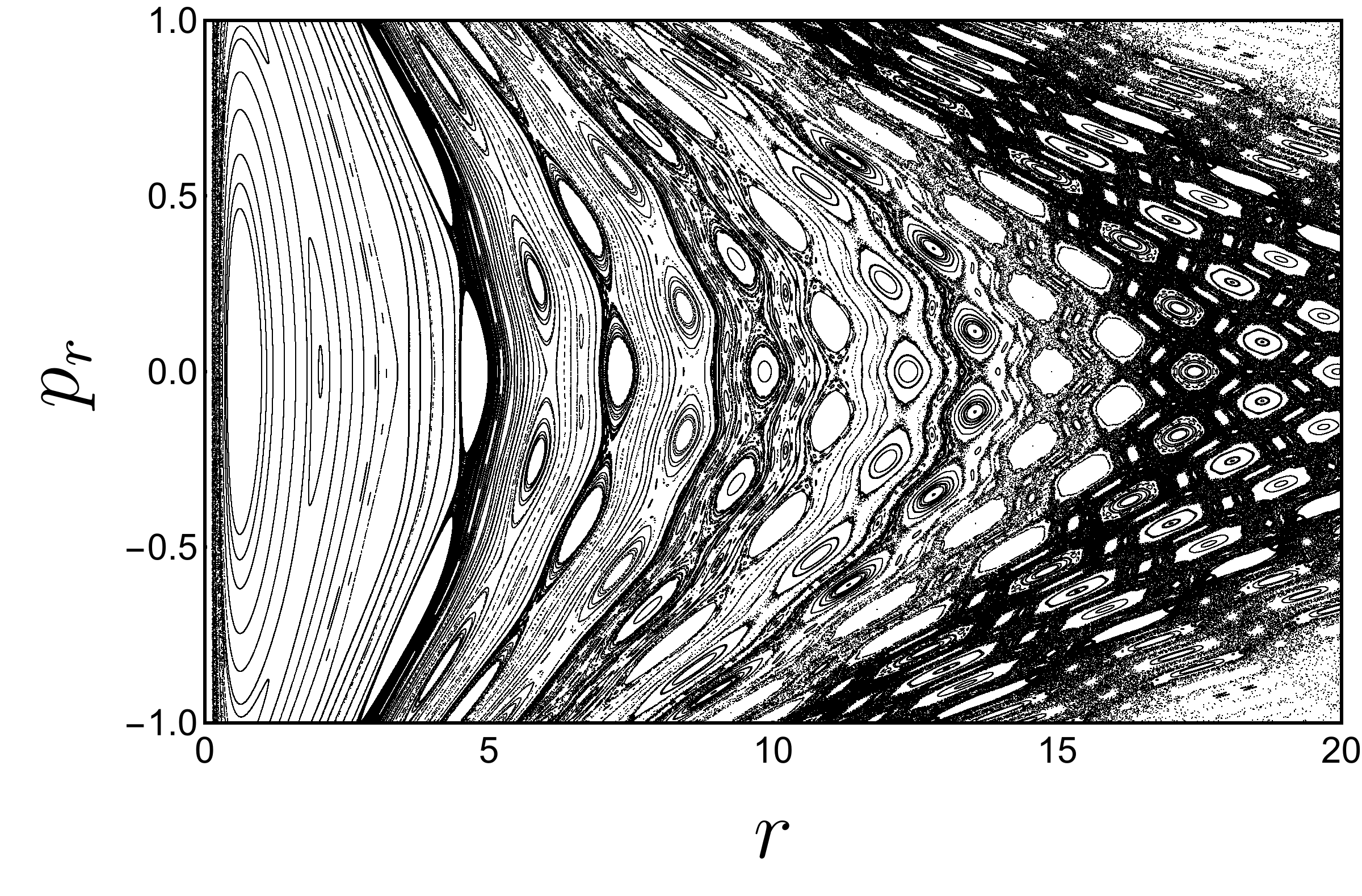}
	\caption{Top, left: Same as figure~\ref{fig:Plummer} but for the logarithmic dark halo potential (\ref{eq:LogPot}), in units where $v_0=1$, $a=1$. We fix $L^2=0.1$. There are no escaping orbits, since $\Phi\to\infty$ as $r\to\infty$.
		Top, right: $q_0/m=0.001$ and $\Omega=2.0$. 
		Bottom, left: $q_0/m=0.005$ and $\Omega=2.0$. 
		Bottom, right: $q_0/m=0.01$ and $\Omega=2.0$.	
	}
	\label{fig:Log}
\end{figure*}

\section{Dynamics in dark matter halo fields}\label{sec:DMH}

We now consider the orbital dynamics of pulsating spherical objects in the presence of diverse spherically symmetric dark matter halos. 
In this case, motion remains restricted to the equatorial plane of the system (even if with a time-varying inertia parameter $q(t)$) and the total specific angular momentum $L$ of the body with respect to the origin is conserved  \cite{bolinha}. 

We can thus consider an effective potential formulation for a spherical body whose energy function is given by
\begin{equation}\label{eq:energyCentral}
E=E(r,p_r, t)= \frac{1}{2}\left(p_r\right)^2 + \mathcal{V}_{\rm eff}(r,t)\,,
\end{equation}
where $p_r$ is the radial component of $\bm p$ and
\begin{equation}\label{eq:VVeffball}
\mathcal{V}_{\rm eff}(r,t) = \Phi(r) + \frac{L^2}{2 r^2} + \frac{1}{2m}\,q(t)\,\nabla^2\,\Phi(r).
\end{equation}
It is important to note that the inertia parameter $q(t)$ couples to the Laplacian of the gravitational potential in equation~(\ref{eq:VVeffball}). Therefore, a spherical object (whether pulsating or not) will follow the same trajectory as that of a point particle with the same initial conditions if immersed in a gravitational vacuum. However, in a nontrivial background---with $\nabla^2\,\Phi\neq 0$, as in a dark matter halo---this coupling may lead not only to small corrections to the point-particle trajectory but even to chaotic orbital motion.
In fact, as we show in the following, the presence of dark matter halos gives rise to resonant stability islands and chaotic regions in phase space when pulsations are taken into account in the orbital dynamics of the objects.

In order to explore the chaotic regions in phase space, we construct $(r-p_r)$ stroboscopic Poincar\'e maps given by $t_n=2n\pi/\Omega$ (see equation~(\ref{eq:qt})). These maps are symplectic, since they correspond to Poincar\'e surfaces of section in extended phase space \cite{tabor1989chaos, lichtenbergLieberman1992}. 
We present below numerical results for stroboscopic Poincar\'e sections in the Plummer halo, in the logarithmic halo potential, and in the Navarro-Frenk-White (NFW) and Hernquist profiles.

We typically use in these simulations values of $q_0/m$ of $10^{-3}$ to $10^{-4}$ and $\Omega$ of the order of unity. These were chosen so as to show the appearance of chaos in the several models we discuss. However, these are not typical values for realistic galactic scenarios. 
Let us fix our units such that $R_{\rm gal}\sim 1$ for the characteristic radius of the galaxy. Therefore for a star with radius $R_{\rm star}$ of 100 solar radii, we would have in our units (for a galaxy of 10 kpc of radius) $q_0/m \propto(R_{\rm star}/R_{\rm gal})^2\sim 10^{-20}$. Also, in our units, the period of revolution around the galaxy, $T_{\rm rev}$, is of the order of unity. Since stars have revolution periods of about 100 Myr and stellar oscillations have periods $T_{\rm osc}$ of 1 day -- 10 years, then by the same arguments above we should have in our units $\Omega\sim (T_{\rm rev}/T_{\rm osc})\sim 10^{7}$  at best.
In this way the numerical experiments that follow have the intent of showing that chaos arises in principle; nevertheless it should not be detectable in real galactic systems.

\begin{figure*}
	\includegraphics[width=0.9\columnwidth]{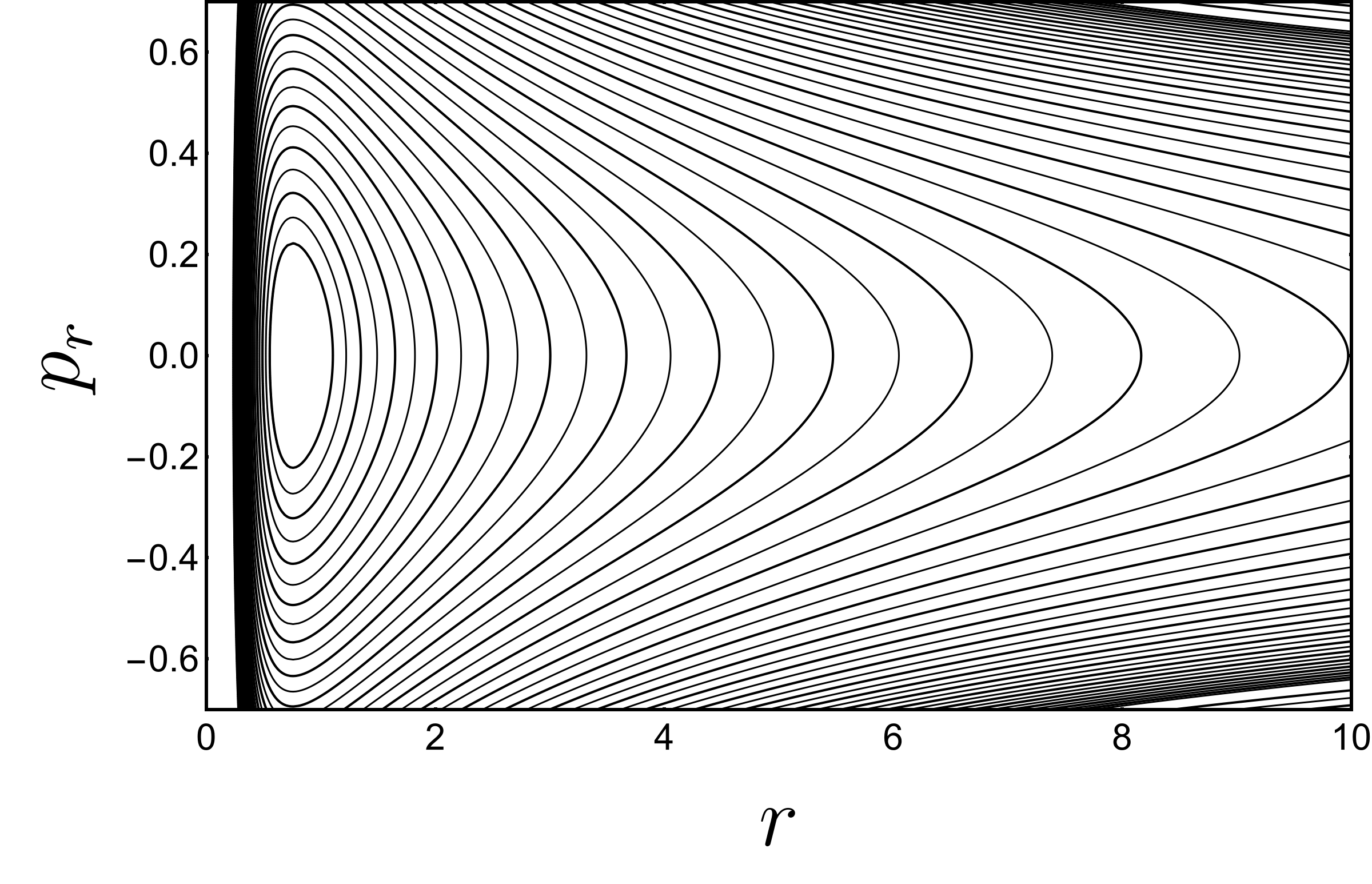}\qquad\qquad
	\includegraphics[width=0.9\columnwidth]{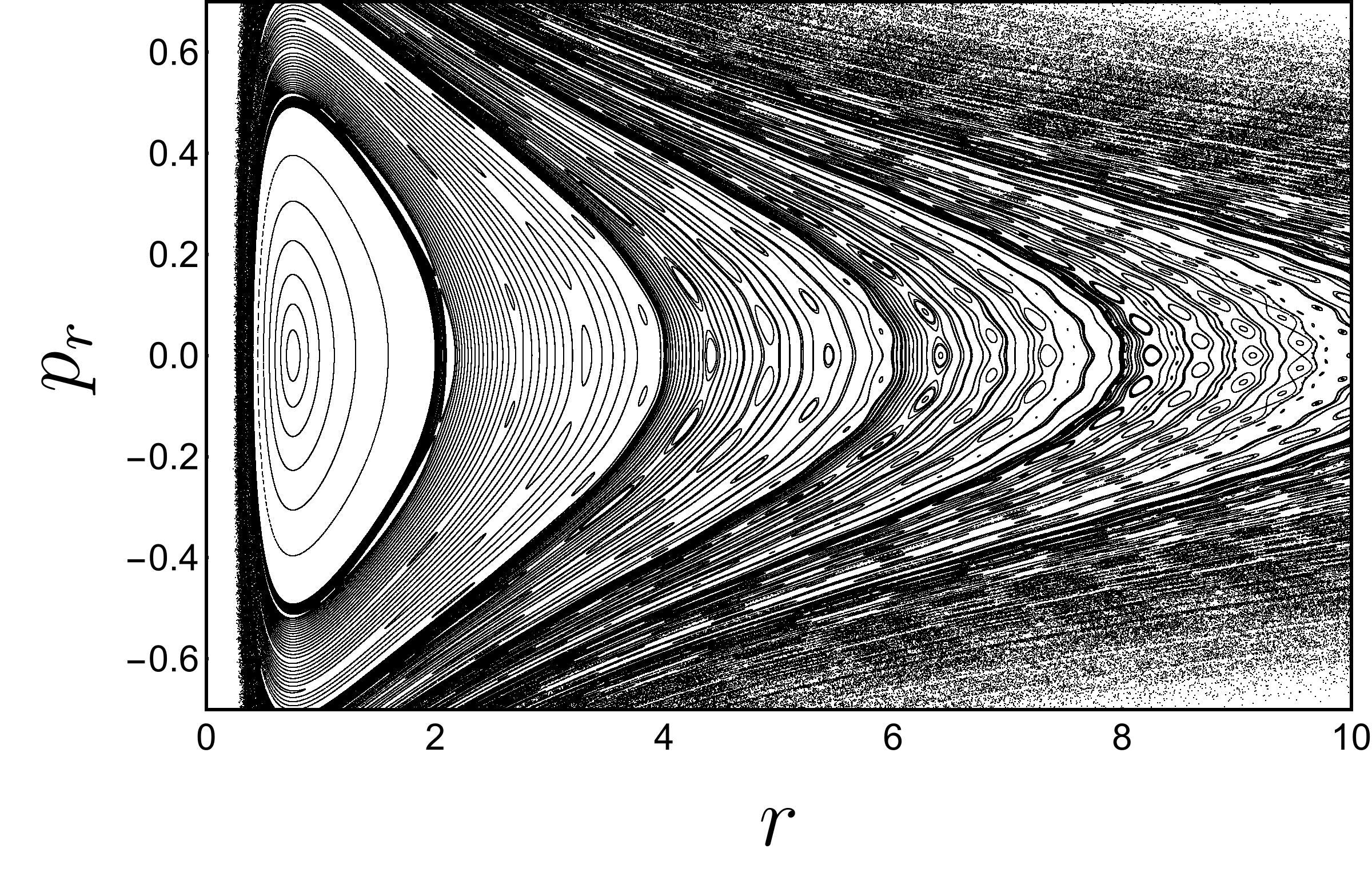}
	\includegraphics[width=0.9\columnwidth]{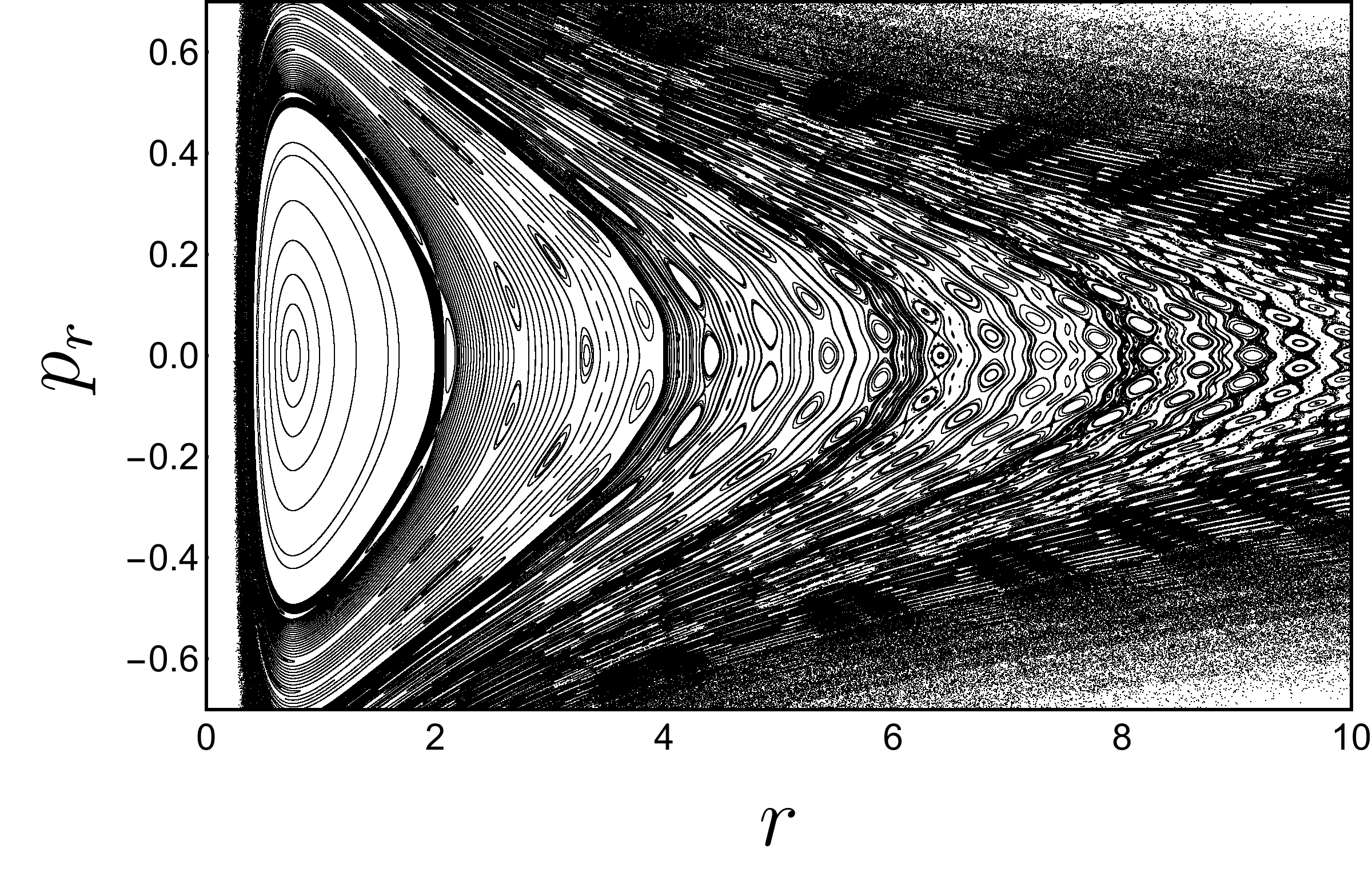}\qquad\qquad
	\includegraphics[width=0.9\columnwidth]{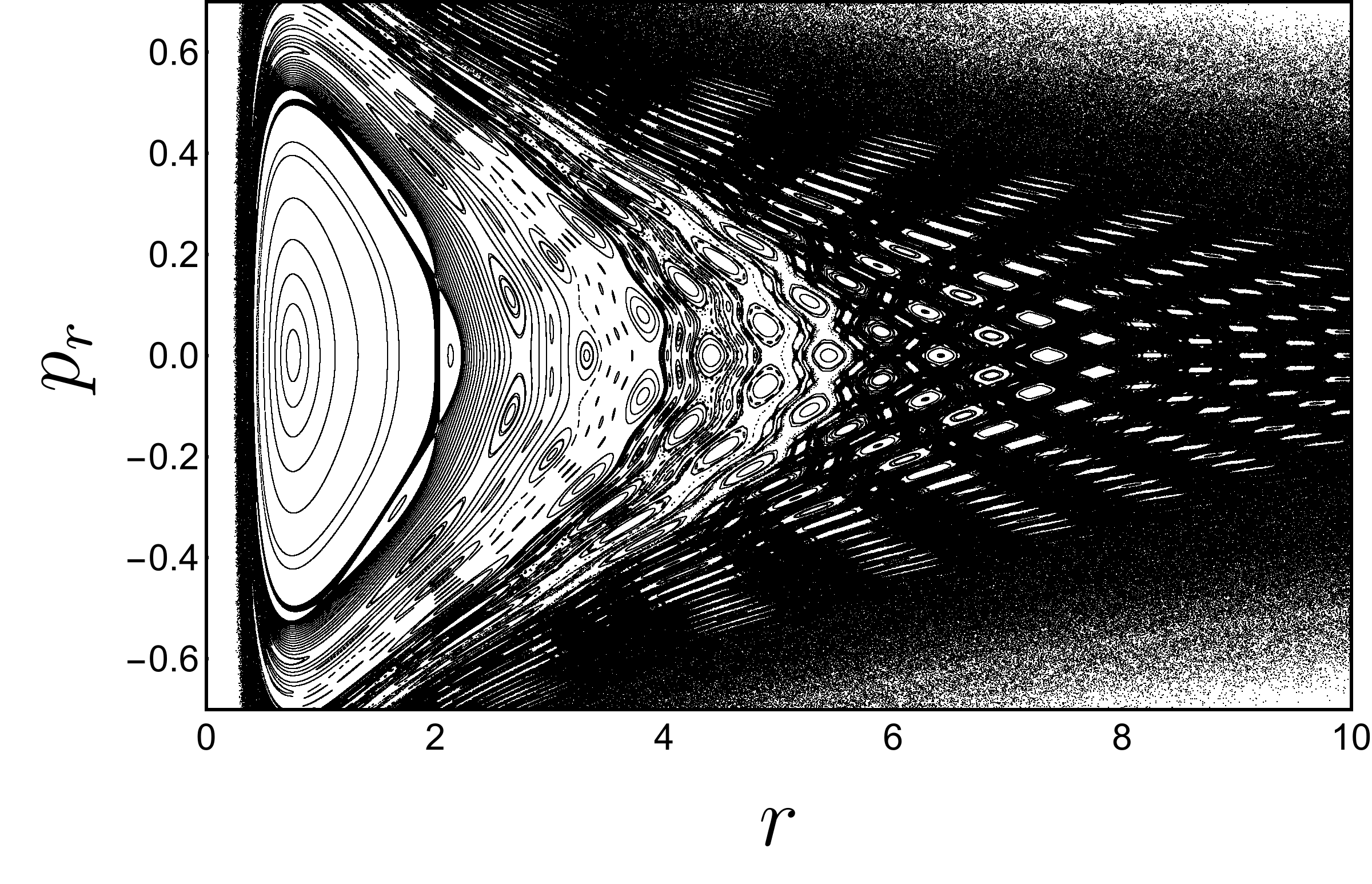}
	\caption{Top, left: Same as figure~\ref{fig:Plummer} but for the NFW dark halo potential (\ref{eq:NFWPot}), in units where $k=1$, $a=1$. We fix $L^2=0.1$. The blank regions correspond to orbits escaping to infinity.
		Top, right: $q_0/m=0.0005$ and $\Omega=2.0$. 
		Bottom, left: $q_0/m=0.001$ and $\Omega=2.0$. 
		Bottom, right: $q_0/m=0.003$ and $\Omega=2.0$.
	}
	\label{fig:NFW}
\end{figure*}

\subsection{Plummer halo}

We first consider a toy model for the dark matter halo given by a Plummer sphere with gravitational potential \cite{binneytremaineGD}
\begin{equation}\label{eq:PlummerPot}
\Phi(r) = -\frac{GM}{\sqrt{r^2+b^2}}.
\end{equation}
This model has been used in the literature by various authors in order to analyse qualitative properties of the dark matter halo field (e.g., \cite{selwoodMcgaugh2005ApJ, vieiraRamoscaro2014ApJ, bobylevEtal2017AstL}).

Figure~\ref{fig:Plummer} shows the phase portrait of orbits in the point-particle approximation as well as stroboscopic ($r-p_r$) Poincar\'e sections for different values of the perturbation amplitude $q_0/m$ (see equation~\ref{eq:qt}). The results presented here are representative of the numerical experiments performed.
We clearly see that pulsation breaks the integrability of the system, and that the chaotic region in phase space gets larger as $q_0/m$ increases.

\subsection{Spherical logarithmic halo}

The logarithmic dark matter halo, firstly proposed by Binney \cite{binney1981MNRAS}, is a theoretical construct which produces an almost flat, constant circular speed rotation curve \cite{binneytremaineGD}. Although its total mass is infinite, the corresponding nonspherical model has proven to be an interesting framework to analyse the overall dark matter halo shape in the Milky Way (e.g., \cite{helmi2004MNRAS, ruzickaEtal2007AA, lawEtal2009ApJL, degWidrow2013MNRAS}). Its potential is given by \cite{binneytremaineGD}
\begin{equation}\label{eq:LogPot}
\Phi(r) = \frac{1}{2} v_0^2\,\log\left(1+\frac{r^2}{a^2}\right),
\end{equation}
where $v_0$ is the corresponding asymptotic circular speed.

Figure~\ref{fig:Log} shows stroboscopic Poincar\'e maps for different values of $q_0/m$, also implying the spread of chaotic regions and resonance chains as $q_0/m$ increases.

\subsection{Navarro-Frenk-White halo}

The NFW halo appears as a `universal dark matter density profile' from cosmological $N$-body simulations of structure formation in the Universe \cite{navarroFrenkWhite1996ApJ, navarroFrenkWhite1997ApJ}. It is therefore of great interest to verify whether chaos is also present in this realistic model. Its gravitational potential is given by \cite{navarroFrenkWhite1996ApJ, navarroFrenkWhite1997ApJ, binneytremaineGD}
\begin{equation}\label{eq:NFWPot}
\Phi(r) = -\frac{k}{r/a}\, \ln\left(1+\frac{r}{a}\right),
\end{equation}
where $k = 4\pi G \rho_0 a^2$.
The calculation of stroboscopic Poincar\'e sections indeed shows the appearance of chaotic regions surrounding islands of stability when the perturbation parameter $q_0/m$ is taken into account, as exemplified in figure~\ref{fig:NFW}. 

\begin{figure*}
	\includegraphics[width=0.9\columnwidth]{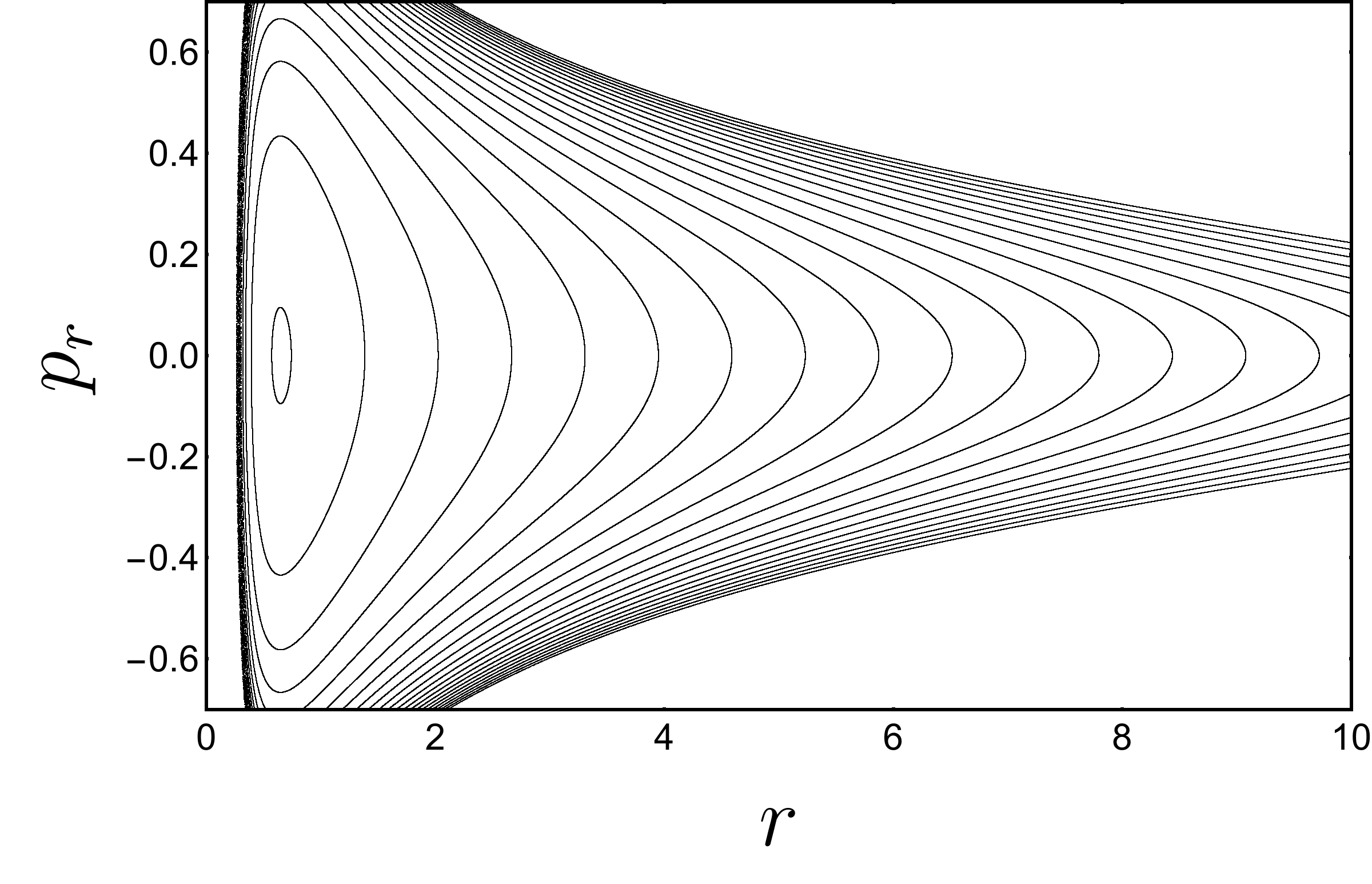}\qquad\qquad
	\includegraphics[width=0.9\columnwidth]{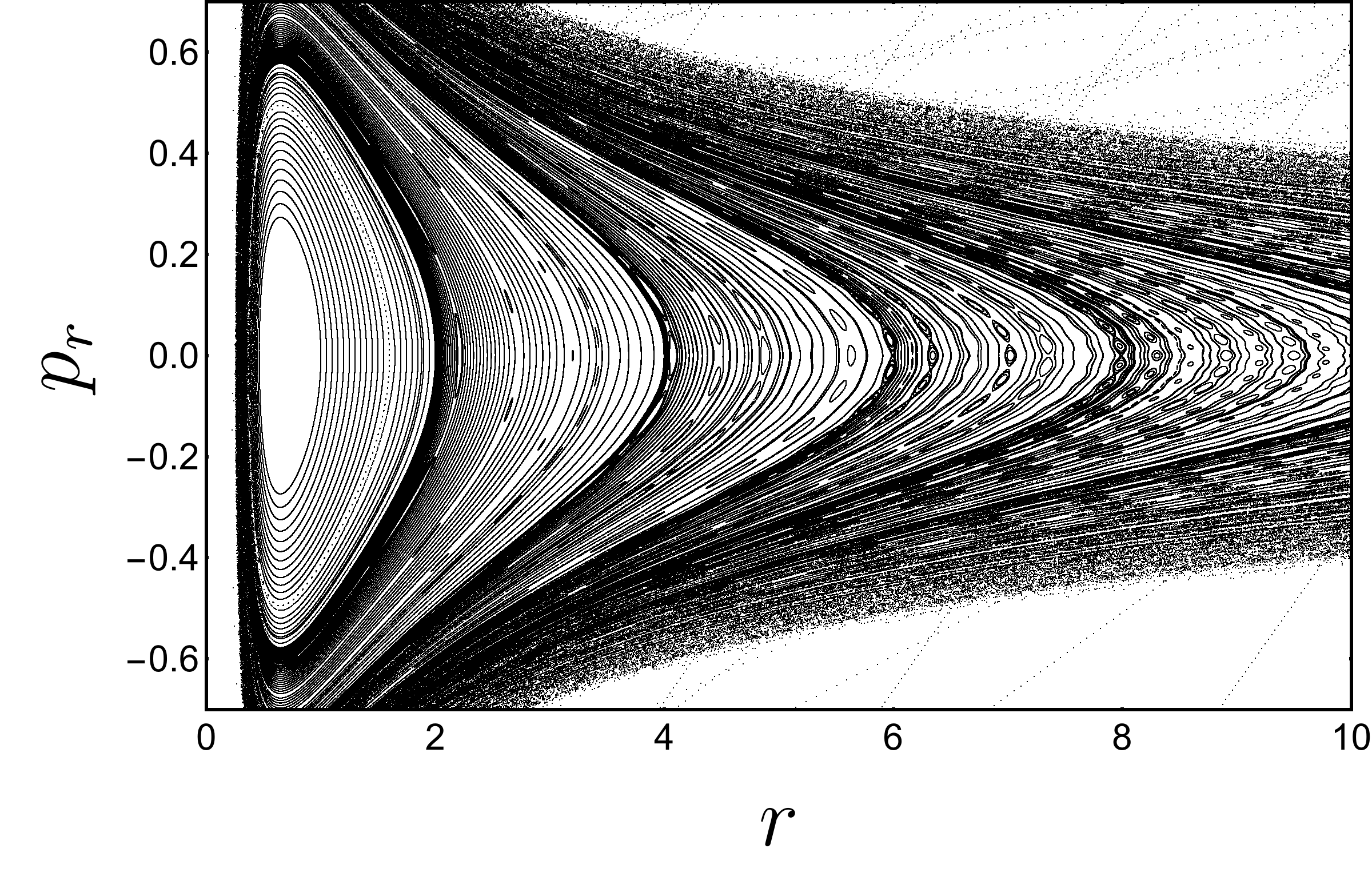}
	\includegraphics[width=0.9\columnwidth]{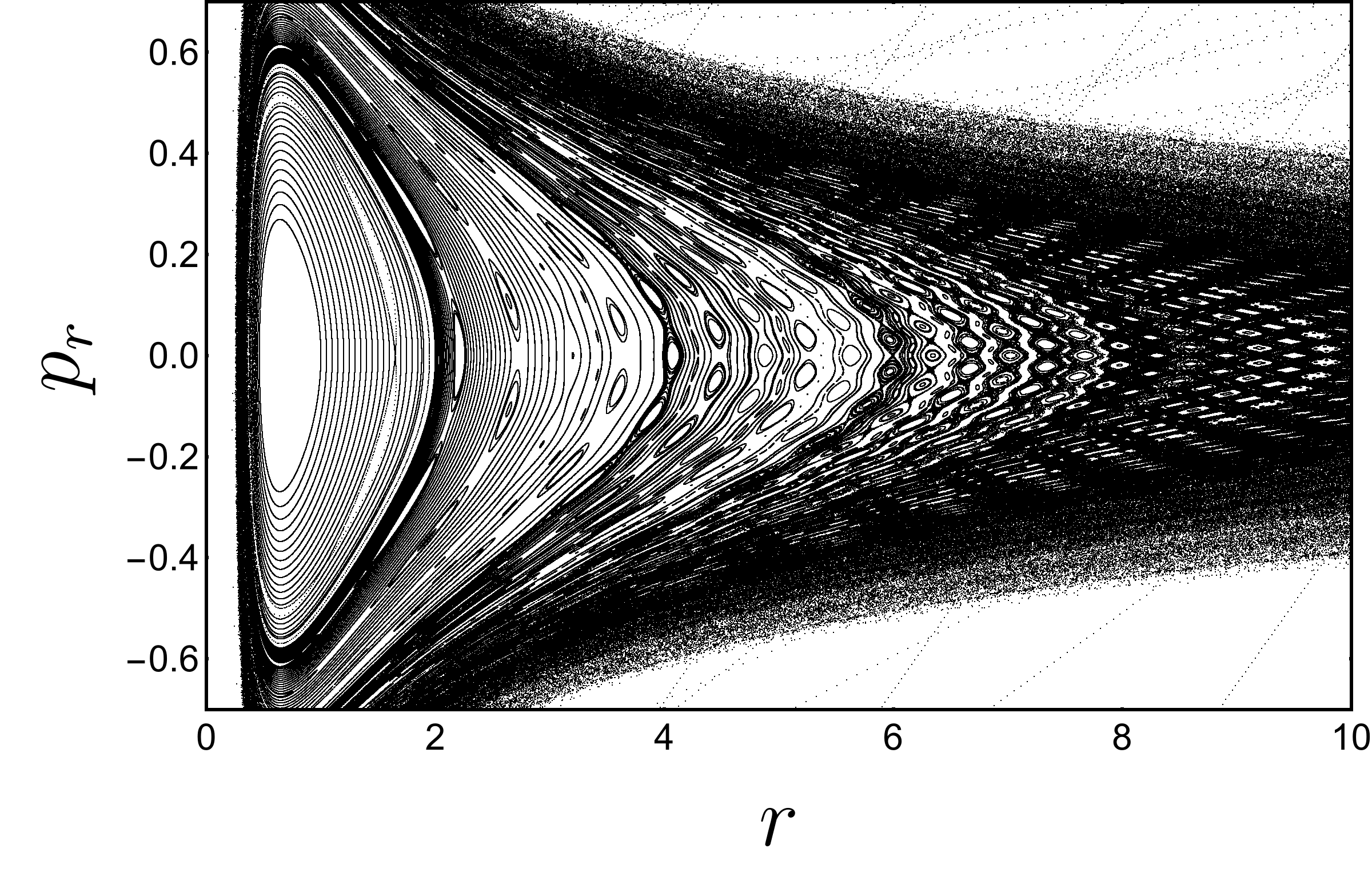}\qquad\qquad
	\includegraphics[width=0.9\columnwidth]{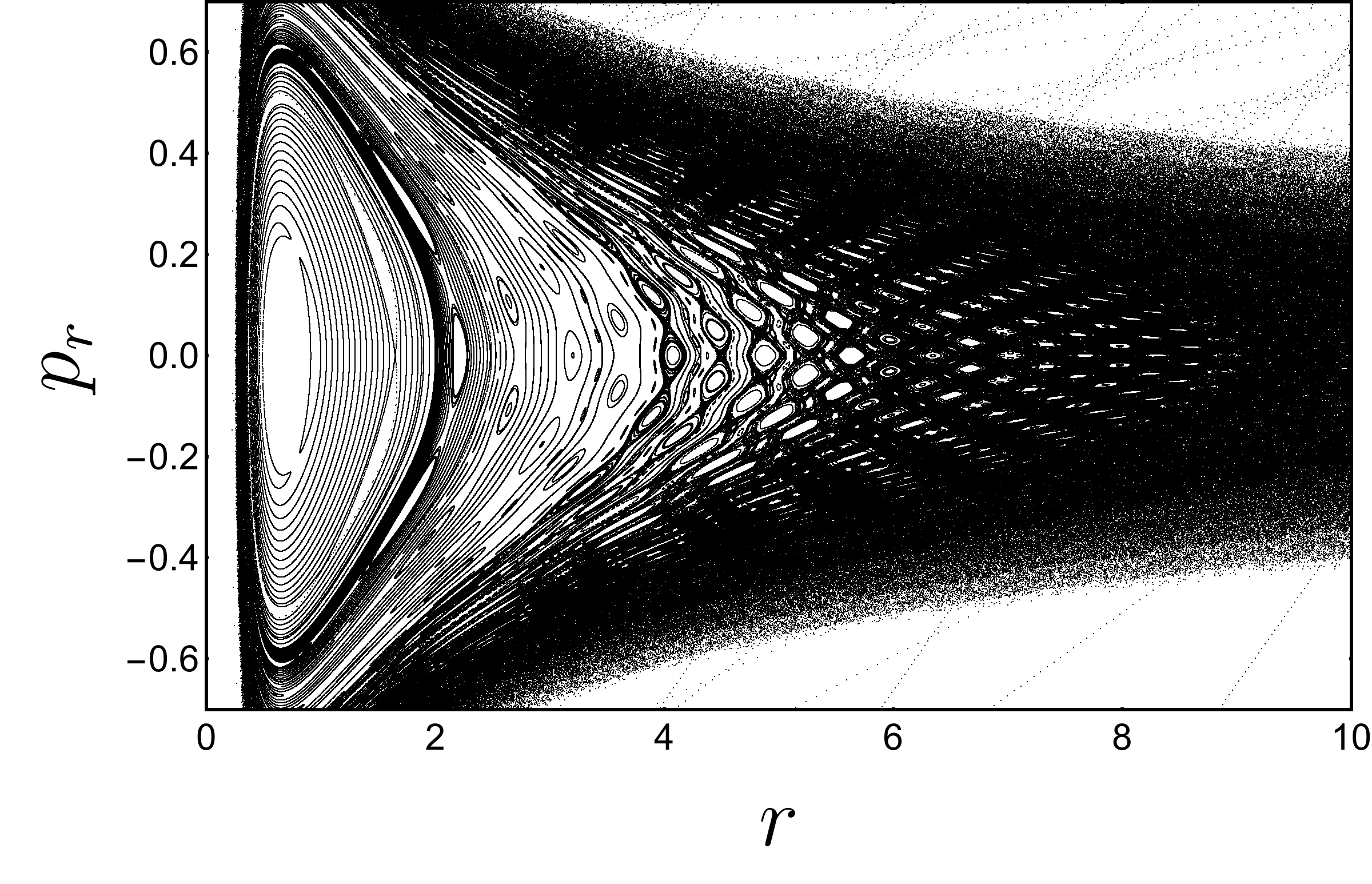}
	\caption{Top, left: Same as figure~\ref{fig:Plummer} but for the Hernquist dark halo potential (\ref{HernquistPot}), in units where $k=1$, $a=1$. We fix $L^2=0.1$. The blank regions correspond to orbits escaping to infinity.
		Top, right: $q_0/m=0.0001$ and $\Omega=2.0$. 
		Bottom, left: $q_0/m=0.0005$ and $\Omega=2.0$. 
		Bottom, right: $q_0/m=0.001$ and $\Omega=2.0$.
	}
	\label{fig:Hernquist}
\end{figure*}

\subsection{Hernquist halo}

Finally, we consider a Hernquist halo \cite{hernquist1990ApJ, binneytremaineGD}, whose gravitational potential is given by
\begin{equation}\label{HernquistPot}
\Phi(r) = -\frac{k}{2(1+r/a)},
\end{equation}
where $k=4\pi G \rho_0 a^2$.
Figure~\ref{fig:Hernquist} shows the phase portrait of test-particle motion and the corresponding stroboscopic Poincar\'e sections for different values of $q_0/m$. The overall picture is the same as in the previous cases; chaos appears in the system as $q_0/m$ gets larger.

Therefore, from the above results, the appearance of chaotic regions in the Poincar\'e sections corroborates the assertion that orbital chaos is a typical phenomenon for pulsating objects in the presence of matter. In all cases, the chaotic regions are more prominent for large radii.

\section{Conclusions and final remarks}
\label{sec:conclusions}

We showed that pulsation generates orbital chaos in the presence of a background matter field. We argue that the appearance of chaos in the orbital dynamics of pulsating objects is a generic phenomenon when they are immersed in matter density distributions (that is, in the non-vacuum case). Although we have analysed toy models of pulsating stars immersed in dark matter halos, the same formalism can be applied to pulsating objects in galaxy bulges or in elliptical galaxies
(both usually modelled by a Hernquist profile, whose chaotic dynamics is presented here), or even moving along the mid-plane of axisymmetric disc models. It is worth noting that, since the  inertia parameter $q(t)$ couples to the field's Laplacian by equation~(\ref{eq:potentialBall}), its dynamics on the equatorial plane will be affected by a possible oblateness of the halo (or, in the case of disc potentials, by the vertical structure of the disc near $z=0$), even if the gravitational potential restricted to the equatorial plane is the same. In particular, for (axisymmetric models of) disc galaxies with the same rotation curve, the fate of these orbits may greatly depend on the disc's vertical density distribution. This scenario is qualitatively different from the point-particle approximation, in which the mid-plane orbital dynamics depends only on the rotation curve \cite{binneytremaineGD, michtchenkoVieiraBarrosLepine2017AA}.


The case analysed here (the fundamental radial mode of oscillation) is not the most general perturbation to point-particle motion. As shown in \cite{bolinha}, periodic deviations from spherical symmetry of the body also generate chaotic orbital behaviour, even in Kepler's problem (it also happens in the relativistic regime of Schwarzschild spacetime, see \cite{mosnaRodriguesVieira2022PRD}). Therefore the nonradial oscillations of variable stars might also, in principle, generate chaos in their orbital dynamics. Also, nonlinear, irregular pulsations may generate a much more abrupt route to chaotic orbital motion.
However, we remark that the parameters used in the figures to show chaoticity of orbital motion, when redimensionalised to compare with realistic scenarios, lie far outside the estimated range for typical stars in average-size galaxies. Therefore, although this effect is probably not detectable in the orbital motion of pulsating stars, it is formally present in the corresponding dynamical system when modelled by the prescriptions used here.

\section*{Acknowledgments}

We thank an anonymous referee for pointing out the correct order-of-magnitude estimates for the effect presented in this paper in realistic galactic scenarios.
R.A.M. was partially supported by Conselho Nacional de Desenvolvimento Cient\'{i}fico e Tecnol\'{o}gico under grant 310403/2019-7.






\begin{thebibliography}{1}
	
	
	\bibitem{carroll2006introduction}
	B.~W. Carroll and D.~A. Ostlie,
	\newblock {\em An introduction to modern astrophysics} (Addison-Wesley, San
	Francisco, US, 2006).
	
	\bibitem{cox1974RepPPhys}
	J.~P. Cox,
	\newblock Pulsating stars, Reports on Progress in Physics {\bf 37}, 563 (1974).
	
	\bibitem{cox1980TSP}
	J.~P. {Cox},
	\newblock {\em {Theory of stellar pulsation}} (Princeton University Press,
	Princeton, NJ, 1980).
	
	\bibitem{takeuti2012nonlinear}
	M.~Takeuti and J.~R. Buchler,
	\newblock {\em Nonlinear Phenomena in Stellar Variability} (Springer Science \&
	Business Media, 2012).
	
	\bibitem{sofueRubin2001ARAA}
	Y.~{Sofue} and V.~{Rubin},
	\newblock {Rotation Curves of Spiral Galaxies}, \araa \ {\bf 39}, 137 (2001),
	astro-ph/0010594.
	
	\bibitem{einasto2009arXiv}
	J.~{Einasto},
	\newblock {Dark Matter}, arXiv:0901.0632  (2009).
	
	\bibitem{sanders2010darkmatter}
	R.~H. Sanders,
	\newblock {\em The dark matter problem: a historical perspective} (Cambridge
	University Press, Cambridge, UK, 2010).
	
	\bibitem{cloweEtal2006ApJ}
	D.~{Clowe} {\em et~al.},
	\newblock {A Direct Empirical Proof of the Existence of Dark Matter}, \apjl \
	{\bf 648}, L109 (2006), astro-ph/0608407.
	
	\bibitem{bartelmann2010RvMP}
	M.~{Bartelmann},
	\newblock {The dark Universe}, Reviews of Modern Physics {\bf 82}, 331 (2010),
	0906.5036.
	
	\bibitem{bolinha}
	R.~S.~S. {Vieira} and R.~A. {Mosna},
	\newblock {Homoclinic chaos in the Hamiltonian dynamics of extended test
		bodies}, arXiv:2207.01594  (2022).
	
	\bibitem{harte2021AcAau}
	A.~I. {Harte} and M.~T. {Gaffney},
	\newblock {Extended-body effects and rocket-free orbital maneuvering}, Acta
	Astronautica {\bf 178}, 625 (2021), 2002.10922.
	
	\bibitem{tabor1989chaos}
	M.~Tabor,
	\newblock {\em Chaos and integrability in nonlinear dynamics} (John Wiley \&
	Sons, New York, US, 1989).
	
	\bibitem{lichtenbergLieberman1992}
	A.~Lichtenberg and M.~Lieberman,
	\newblock {\em Regular and Chaotic Dynamics}, Applied Mathematical Sciences
	Vol.~38 (Springer, New York, US, 1992).
	
	\bibitem{binneytremaineGD}
	J.~{Binney} and S.~{Tremaine},
	\newblock {\em Galactic Dynamics}, Second ed. (Princeton Univ. Press,
	Princeton, NJ, 2008).
	
	\bibitem{selwoodMcgaugh2005ApJ}
	J.~A. {Sellwood} and S.~S. {McGaugh},
	\newblock {The Compression of Dark Matter Halos by Baryonic Infall}, \apj \ {\bf
		634}, 70 (2005), astro-ph/0507589.
	
	\bibitem{vieiraRamoscaro2014ApJ}
	R.~S.~S. {Vieira} and J.~{Ramos-Caro},
	\newblock {A Simple Formula for the Third Integral of Motion of Disk-Crossing
		Stars in the Galaxy}, \apj \ {\bf 786}, 27 (2014), 1305.7078.
	
	\bibitem{bobylevEtal2017AstL}
	V.~V. {Bobylev}, A.~T. {Bajkova}, and A.~O. {Gromov},
	\newblock {Refinement of the parameters of three selected model Galactic
		potentials based on the velocities of objects at distances up to 200 kpc},
	Astronomy Letters {\bf 43}, 241 (2017), 1703.01413.
	
	\bibitem{binney1981MNRAS}
	J.~{Binney},
	\newblock {Resonant excitation of motion perpendicular to galactic planes.},
	\mnras \ {\bf 196}, 455 (1981).
	
	\bibitem{helmi2004MNRAS}
	A.~{Helmi},
	\newblock {Is the dark halo of our Galaxy spherical?}, \mnras \ {\bf 351}, 643
	(2004), astro-ph/0309579.
	
	\bibitem{ruzickaEtal2007AA}
	A.~{R{\r{u}}{\v{z}}i{\v{c}}ka}, J.~{Palou{\v{s}}}, and C.~{Theis},
	\newblock {Is the dark matter halo of the Milky Way flattened?}, \aap \ {\bf
		461}, 155 (2007), astro-ph/0608175.
	
	\bibitem{lawEtal2009ApJL}
	D.~R. {Law}, S.~R. {Majewski}, and K.~V. {Johnston},
	\newblock {Evidence for a Triaxial Milky Way Dark Matter Halo from the
		Sagittarius Stellar Tidal Stream}, \apjl \ {\bf 703}, L67 (2009), 0908.3187.
	
	\bibitem{degWidrow2013MNRAS}
	N.~{Deg} and L.~{Widrow},
	\newblock {The Sagittarius stream and halo triaxiality}, \mnras  \ {\bf 428}, 912
	(2013), 1209.6614.
	
	\bibitem{navarroFrenkWhite1996ApJ}
	J.~F. {Navarro}, C.~S. {Frenk}, and S.~D.~M. {White},
	\newblock {The Structure of Cold Dark Matter Halos}, \apj \ {\bf 462}, 563
	(1996), astro-ph/9508025.
	
	\bibitem{navarroFrenkWhite1997ApJ}
	J.~F. {Navarro}, C.~S. {Frenk}, and S.~D.~M. {White},
	\newblock {A Universal Density Profile from Hierarchical Clustering}, \apj {\bf
		490}, 493 (1997), astro-ph/9611107.
	
	\bibitem{hernquist1990ApJ}
	L.~{Hernquist},
	\newblock {An analytical model for spherical galaxies and bulges}, \apj \ {\bf
		356}, 359 (1990).
	
	\bibitem{michtchenkoVieiraBarrosLepine2017AA}
	T.~A. {Michtchenko}, R.~S.~S. {Vieira}, D.~A. {Barros}, and J.~R.~D.
	{L{\'e}pine},
	\newblock {Modelling resonances and orbital chaos in disk galaxies. Application
		to a Milky Way spiral model}, \aap \ {\bf 597}, A39 (2017), 1608.08991.
	
	\bibitem{mosnaRodriguesVieira2022PRD}
	R.~A. Mosna, F.~F. Rodrigues, and R.~S.~S. Vieira,
	\newblock Chaotic dynamics of a spinless axisymmetric extended body around a
	schwarzschild black hole, Phys. Rev. D {\bf 106}, 024016 (2022).
	
\end{thebibliography}








\end{document}